\documentclass[prd,nofootinbib,twocolumn]{revtex4}
\usepackage{graphics}
\usepackage{bm}

\begin{document}

\title{
Topological properties of the Cosmic Microwave Background 
polarization map
}

\author{Tanmay Vachaspati\footnote{txv7@po.cwru.edu} \&
        Arthur Lue\footnote{axl71@po.cwru.edu}}
\affiliation{
Center for Education and Research in Cosmology and Astrophysics,
Department of Physics, Case Western Reserve University,
10900 Euclid Avenue, Cleveland, OH 44106-7079, USA.}

\begin{abstract}
\noindent
Rapid progress has been made in observations of the temperature
anisotropies of the Cosmic Microwave Background (CMB). These
observations have enabled cosmologists to characterize the state of
the universe at recombination, and observational efforts are now being
directed towards obtaining the polarization map of the CMB.  Here we
draw an analogy between the CMB polarization map and nematic liquid
crystals, pointing out that the two have similar defects.  Making use
of known results in the theory of defect formation we predict the
statistical distribution of defects in the CMB polarization map and
provide a novel tool for probing the CMB at large angular scales.  
\end{abstract}

\maketitle


A polarization map of the CMB will consist of headless 
vectors laid on the surface of last scattering. One can 
view the polarization map as rod-shaped molecules laid out 
on a two-dimensional sphere ($S^2$), the sky. This picture 
immediately connects the CMB polarization map with 
nematics where the molecules are also rod-shaped. The 
topological properties of nematics have been studied 
for many decades \cite{Mer79}. The very topological properties 
that are relevant to nematics will also apply to the CMB 
polarization map, as first examined in Ref. \cite{DolDorNovNov99}.

At first sight, it might appear that the analogy is doomed
because neighboring molecules in nematics interact,
while CMB polarization vectors do not. However, the processes
at last scattering that produce the polarization are described
by interacting field theory and so neighboring polarization
vectors, just as neighboring liquid crystal molecules, 
tend to align. The CMB polarization map is just like a
snapshot of a thin film of nematic liquid crystal.

The mathematical description of the CMB polarization 
map\cite{Kos99} in a local patch of the sky is in terms 
of a $2\times 2$ symmetric traceless matrix which we denote 
by ${\bf P}$: 
\begin{equation}
{\bf P} = \pmatrix{Q&U\cr U&-Q}
\label{pol_matrix}
\end{equation}
We define a normalized ``order parameter'' as:
${\bf M} = {1\over 2}{\bf P}/\sqrt{Q^2+U^2}$.
The matrix ${\bf M}$ can also be written in terms of a
unit vector ${\hat n}$ as:
\begin{equation}
{\bf M}_{ij} = {\hat n}_i {\hat n}_j - {1\over 2}\delta_{ij}
\label{m-and-n}
\end{equation}
So the space of all matrices ${\bf M}$ is given by the
space ($S^1$) of all two-dimensional unit vectors but with 
the $Z_2$ identification under ${\hat n} \rightarrow -{\hat n}$.
The normalized field ${\bf M}$ therefore defines a unit vector 
field with no arrows, also known as a ``line field.''

Since ${\bf M}$ lies in $S^1/Z_2$, a space that has incontractable
loops, the line field will have singularities, also called
``defects.''  Some of these singularities are shown in 
Fig.~1. The radial and tangential singularities 
commonly discussed in the CMB literature and shown in
Fig.~2 can be constructed by combining two of the 
fundamental singularities.

\begin{figure}
\scalebox{0.40}{\includegraphics{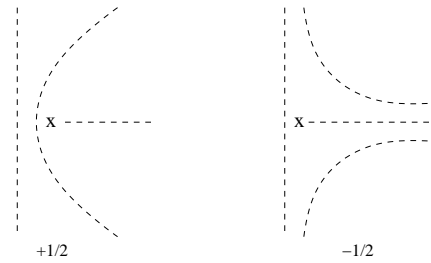}}
\caption{\label{fund_defects}
Fundamental defects of charge $\pm 1/2$. Each dash
respresents the linear polarization of the CMBR at that
point. The $\times$ marks the position of the singularity
where the linear polarization vanishes for topological
reasons.
}
\end{figure}

\begin{figure}
\scalebox{0.40}{\includegraphics{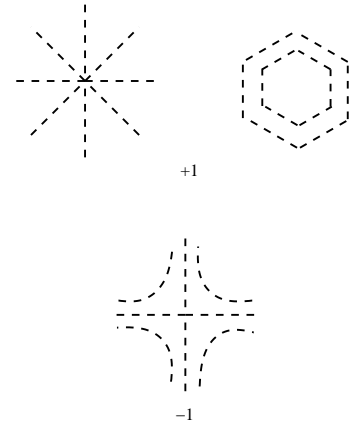}}
\caption{\label{n=1}
Defects of charge $\pm 1$. These can be constructed by
combining two fundamental defects shown in Fig.~1.
}
\end{figure}

In Euclidean space, each singularity has a degree $W$ which 
is given by:
\begin{equation}
W_\Gamma = {1\over {2\pi}} \oint_\Gamma dx^i ~
              {\rm Tr}({\bf M} {\bf \epsilon} \partial_i {\bf M})
\label{charge}
\end{equation}
where $\Gamma$ is a contour around the location of
the singularity, and
\begin{equation}
{\bf \epsilon} \equiv \pmatrix{0&1\cr -1&0}
\label{epsilon}
\end{equation}
In terms of the functions $Q$ and $U$, we simply have
\begin{equation}
W_\Gamma = {1\over {2\pi}} \oint_\Gamma dx^i \partial_i \alpha
\label{qQU}
\end{equation}
where $\alpha = (1/2){\rm tan}^{-1}(U/Q)$. The fundamental defects
shown in Fig. 1 have degrees $\pm 1/2$, while the defects in
Fig. 2 have degrees $\pm 1$. In general we can take $\Gamma$
to be any closed contour on which the polarization is everywhere
non-vanishing, and then Eq. (\ref{qQU}) tells us the
number of singularities within that contour. (We
assume that all the singularities are isolated.)

Equation (\ref{charge}) can also be used on
the surface of a sphere provided we first parallel transport
${\bf M}$ along a geodesic from points on $\Gamma$ to a point 
within $\Gamma$. In other words, to compute the degree, we have
to parallel transport ${\bf M}$ to a common tangent space.

The Poincare-Hopf theorem \cite{ChoDeW89} tells us that
the algebraic sum of degrees of all singularities, also
called the ``index'' of a vector field, on a compact manifold 
is equal to the Euler number of the manifold. 
By explicit evaluation we can show that the
Poincare-Hopf result also applies to a {\em line} field on $S^2$.
Hence the CMB polarization map must necessarily have index 
2 since the Euler number of $S^2$ is 2. 

The distribution of defects produced at a phase transition 
has received much attention over the past few decades 
\cite{Kib76}. A key 
result is that if we consider the length, $L$, of $\Gamma$ 
to be much larger than the correlation scale, $\xi$, of the 
matrix ${\bf M}$, then we can write \cite{VacVil84}
\begin{equation}
W_{\Gamma, rms} \equiv
\langle ~ W_\Gamma^2 ~ \rangle^{1/2} = 
                       c \left ( {L \over \xi} \right )^\beta
\label{qGamma}
\end{equation}
where $c$ is a system-dependent constant but the exponent
$\beta$ is universal with the value $\beta = 1/2$. The
reasoning leading to this scaling is very simple. $W_\Gamma$
is the sum of random fluctuations in $\alpha$ and the
number of elements in the sum goes like $L/\xi$. Therefore
the root mean square value of $W_\Gamma$ is proportional to
$\sqrt{L/\xi}$.

Satellite observations of the CMB \cite{wmap,cobe} show that the
temperature anisotropy correlation function, $C(\theta)$, falls off
very steeply with increasing angular separation, and becomes very
small above an angular scale $\theta = \chi \sim$ a few degrees. 
If this lack of large-scale temperature correlations is indeed 
real, and not some misunderstood systematic effect, it is natural 
to expect that one should find a corroborating lack of correlation 
for polarization on all angular scales larger than $\chi$.  After 
all, thermal and polarization
fluctuations are expected to arise from similar processes during
recombination.  Hence, on angular scales larger than $\chi$, the
matrix ${\bf M}$ would be essentially uncorrelated. This tells us that
the areal density of defects on the sky will be approximately one per
$\chi^2 \sim 10$ square degrees.  Further, we expect $\beta=1/2$ on
angular scales larger than $\chi$.  If the data shows
$\beta \ne 1/2$ or exponential decay with $L$, then that would indicate
the presence of large-scale correlations in the CMB
polarization. This observation would be equivalent to defect formation
with ``bias'' in condensed matter systems.

If the statistical distribution of the polarization is
Gaussian, then all correlation functions of the polarization 
matrix ${\bf P}$ can be calculated in terms of the two point 
correlation function. However, correlators such as 
$\langle W_\Gamma ^2  \rangle$ that are central
to our discussion, cannot be written in terms of two point 
correlators of ${\bf P}$, since they are given by correlators
of the {\em normalized} matrix ${\bf M}$ which involves dividing 
${\bf P}$ by $\sqrt{Q^2+U^2}$. Alternately, 
$\langle W_\Gamma ^2 \rangle$ depends on the two point
correlation function of the angle 
$\alpha = (1/2){\rm tan}^{-1}(U/Q)$ and this is not
given by the two point correlation functions of $U$ and $Q$. 
Hence the information provided by the distribution of defects 
is complementary to that provided by the two point correlators
of the polarization matrix usually discussed. 

There is a global constraint on the CMB polarization map, namely that
``B-modes'' are absent or highly suppressed
\cite{KamKosSte97,SelZal97,KamKosSte97b,SelZal97b} and the reader may
worry that this would affect the distribution of defects. However, the
B-mode relates the variation of the polarization intensity,
$\sqrt{Q^2+U^2}$, to the distribution of the unit vector ${\hat
n}$. The topology, however, does not depend on the polarization 
intensity, only on the direction ${\hat n}$. Hence for a fixed
distribution of ${\hat n}$, we would need to find $\sqrt{Q^2+U^2}$
such that the B-modes vanish. One can show that even pure E-modes
exhibit the defects shown in Figs.~1 and 2 \cite{Bunetal03}.

As polarized CMB radiation propagates from the last scattering
surface to us, intervening effects such as weak lensing
and Faraday rotation can affect the polarization map.
As long as these effects are continuous, they can only move
the locations of the existing defects or, create or destroy
defect-antidefect pairs. This is similar in spirit to the 
coarsening of defects observed in condensed matter systems,
though the actual dynamics of polarization defects is likely
to be very different. 

In conclusion, we have drawn an analogy between the CMB polarization
map and nematic liquid crystals, classified the defects in the 
polarization map, and made a prediction for the scaling of the 
index of the CMB polarization map. Ongoing observations of the 
CMB will be able to find the exponent $\beta$ and compare it to 
the theoretical prediction of $1/2$. Further statistical measures
of the distribution of defects can provide additional novel probes 
of the CMB.

\begin{acknowledgments} 
We thank Craig Copi, Dragan Huterer, Arthur Kosowsky, Harsh Mathur,
Levon Pogosian, John Ruhl, Rafael Sorkin, Glenn Starkman, and Phil
Taylor for discussions.  This work was supported by the
U.S. Department of Energy and the CWRU Office of the Provost.
\end{acknowledgments}

\end{document}